# How to get mechanical work from a capacitor and a couple of batteries

## E. N. Miranda

Área de Ciencias Exactas
CRICYT
5500–Mendoza, Argentina

Facultad de Ingeniería
Universidad de Mendoza
5500-Mendoza, Argentina

Departamento de Física
Universidad Nacional de San Luis
5700-San Luis, Argentina

**Abstract:** The work done by a parallel plate capacitor is evaluated when the plate separation is changed. Two cases are considered: 1) the capacitor has a constant charge; 2) the capacitor is at constant voltage. The net work is calculated when the device follows a closed cycle in the charge-voltage space. For certain conditions a net mechanical work can be obtained from the cycling capacitor. The analysis is simple enough to be explained in a general physics course.

If we think about getting mechanical work from a battery, an electrical engine comes immediately to mind. This is a natural association since it is the usual way of transforming electrochemical energy stored in the battery into mechanical work [1, 2]. However, in this article an alternative way will be shown. It is a novel approach using a capacitor and a couple of batteries with different voltages. The analysis of the system should be understandable by freshman students with a good background in general physics [3, 4].

We will consider in our analysis a parallel plate capacitor to get exact results; however, the general idea should work for any shape. The area of the capacitor plates is $A$, the gap between them, filled with air, is $x$ so the electric permeability of the dielectric medium is $\varepsilon_0$.

Our first step is to study the work done by a capacitor with a fixed stored charge and a variable gap between its plates. The physical situation is the following. A capacitor is initially uncharged and the plate gap is $x_i$. A battery with a voltage $V_i$ is connected to the capacitor, and it gets a charge $q$. Afterwards, it is disconnected from the battery but remains charged; the plate space is changed to $x_f$, and work has to be done on the capacitor in order to modify the gap; our aim is to evaluate that work $W_q$. The subindex $q$ means the system is with constant charge.

To calculate $W$, it is sufficient to evaluate the change in the capacitor potential energy $E$ written as $E = q/2C$ where $C$ is the capacitance. For a parallel plate capacitor, the capacitance is $C = \varepsilon_0 A/x$ and we can write:

$$\begin{aligned} W_q &= E_f - E_i \\ &= \frac{1}{2}\frac{q^2}{C_f} - \frac{1}{2}\frac{q^2}{C_i} \\ &= \frac{1}{2}\frac{q^2}{\varepsilon_0 A}(x_f - x_i) \end{aligned} \tag{1}$$

Eq. (1) explicitly states that the stored charge remains constant in the process. However, it would be useful to restate it in terms of the battery voltage. One should remember that $C = q/V$; consequently, for a plane capacitor one gets:

$$W_q = \frac{\varepsilon_0 A V_i^2}{2 x_i^2}(x_f - x_i) \tag{2}$$

We have found the work done by a parallel plane capacitor that has been connected to a battery with voltage $V_i$, got a charge $q$, disconnected from the battery and its plate separation changed from $x_i$ to xf. Since the process is at constant charge $q = C_i V_i = C_f V_f$ and:

$$\frac{V_i}{x_i} = \frac{V_f}{x_f} \tag{3}$$

This relation will be useful later.

The next step is to evaluate the work done when the voltage is constant. The capacitor is connected to a battery with voltage $V$, and the plate separation is changed from $x_i$ to $x_f$. One should remember that the force among the plates can be written as:

$$\begin{aligned} F &= -\frac{\partial E}{\partial x} \\ &= -\frac{\partial}{\partial x}\left(\frac{1}{2}\frac{\varepsilon_0 A V^2}{x^2}\right) \\ &= \frac{\varepsilon_0 A V^2}{2x} \end{aligned} \tag{4}$$

The work needed to change the plate gap comes to be:

$$\begin{aligned} W_V &= \int_{x_i}^{x_f} F dx \\ &= \int_{x_i}^{x_f} \frac{\varepsilon_0 A V^2}{2x^2} \\ &= \frac{\varepsilon_0 A V^2}{2}\left(\frac{1}{x_i} - \frac{1}{x_f}\right) \end{aligned} \tag{5}$$

The subindex $V$ means the process has taken place at constant voltage.

We have all the elements required to use the capacitor as a device that can deliver mechanical work. The capacitor has to follow the cycle showed in Figure 1. The analogy with thermodynamic cycles of thermal engines leads us to use a similar terminology of "expansions" and "compressions". The capacitor starts at the state A with coordinates ($x_1$, $V_1$) and is expanded at constant charge to B($x_2$, $V_2$); then, it is again expanded at constant voltage from B($x_2$, $V_2$) to C($x_3$, $V_2$). From C, it is compressed at constant charge until it reaches the original voltage in D($x_4$, $V_1$); finally, the cycle is closed with a compression from D($x_4$, $V_1$) to A.

Our task is to evaluate each step to get the net mechanical work done (or received) by the capacitor.

From A to B, the work is at constant charge; then, according to eq. (1):

$$W_{A,B} = \frac{\varepsilon_0 A V_1^2}{2x_1^2}(x_2 - x_1) \tag{6}$$

From B to C, the system is at constant potential; following eq. (5), the work is:

$$W_{B,C} = \frac{\varepsilon_0 A V_2^2}{2}\left(\frac{1}{x_2} - \frac{1}{x_3}\right) \tag{7}$$

From C to D, one gets:

$$W_{C,D} = \frac{\varepsilon_0 A V_1^2}{2 x_3^2}(x_4 - x_3) \tag{8}$$

Finally, from D to A, it is

$$W_{D,A} = \frac{\varepsilon_0 A V_1^2}{2}\left(\frac{1}{x_4} - \frac{1}{x_1}\right) \tag{9}$$

The net work is the sum of equations (6)-(9). It should be noticed that $x_2$ and $x_4$ can be written, according to (3), as: $x_2 = x_1 V_2/V_1$, $x_4 = x_3 V_1/V_2$. The final outcome is:

$$W_{net} = \varepsilon_0 A (V_2 - V_1)\left(\frac{V_1}{x_1} - \frac{V_2}{x_3}\right) \tag{10}$$

Eq. (10) is the central result of this paper and shows that the net work done by the capacitor performing the cycle shown in Figure 1 is different from zero. One may get mechanical work if $V_2/x_3 > V_1/x_1$.

A graphical interpretation of the previous result is possible. The cycle has to be plotted in the $(q, V)$ space as shown in Figure 2. In this space the process plot is a rectangle and the enclosed area is easy to evaluate. One can write:

$$q_1 = C_1 V_1 = \frac{\varepsilon_0 A}{x_1} V_1$$
$$q_1 = C_3 V_2 = \frac{\varepsilon_0 A}{x_3} V_2 \tag{11}$$

Consequently, the rectangle area is:

$$\text{Area} = \varepsilon_0 A \left(\frac{V_2}{x_3} - \frac{V_1}{x_1}\right)(V_2 - V_1) \tag{12}$$

The area enclosed by the rectangle is the net work done by the system with the opposite sign. Our aim is accomplished: we can get net work from a capacitor following the cycle of Figure 1 or 2 and shown by eq. (10) or (12) if the relation between $V_1$, $V_2$, $x_1$ and $x_3$ is the correct one.

It is pointless to evaluate the efficiency of the system analyzed here since all the work delivered by the capacitor comes from the batteries with voltages $V_1$ and $V_2$. It would be more realistic to consider the wire resistance and get the Joule heat dissipated. However, it is difficult to evaluate the current flowing between the batteries and the capacitor because its capacitance changes with the plate separation. One should know the temporal evolution of the gap $x(t)$ to get the time dependence of the capacitance $C(t)$.

To summarize, an engine might be built from two batteries with potentials $V_1$ and $V_2$ and a capacitor. Although of little technological import, it is interesting from a conceptual point of view and affords a useful exercise for a general physics course.

**Acknowledgment:** The author thanks partial financial support from the National Scientific and Technological Research Council of Argentina (CONICET).

# Figure captions:

**Figure 1:** This graph shows the closed cycle followed by the system. The capacitor is connected to a battery with potential $V_1$ at point A and the plate gap is $x_1$. Afterwards, the battery is disconnected and the plate separation is increased, keeping constant the charge in the capacitor. Once the voltage among the plates is $V_2$, the capacitor is connected to a battery with that potential. With the capacitor at a constant voltage, the plate separation is increased again up to $x_3$, reaching point C. Then, the plate gap is decreased until the original voltage $V_1$ is reached at point D. From this point, a new "compression" at constant voltage takes place and the original plate separation is regained. Notice that the slope of AB (CD) is proportional to the charge stored in the capacitor.

**Figure 2:** The same cycle depicted in the previous figure is shown in the (q, V) plane. Since each cycle step is at constant charge or at constant voltage, the cycle shape is rectangular. It is a simple exercise to evaluate the area of the rectangle that comes to be minus the net work performed by the system.

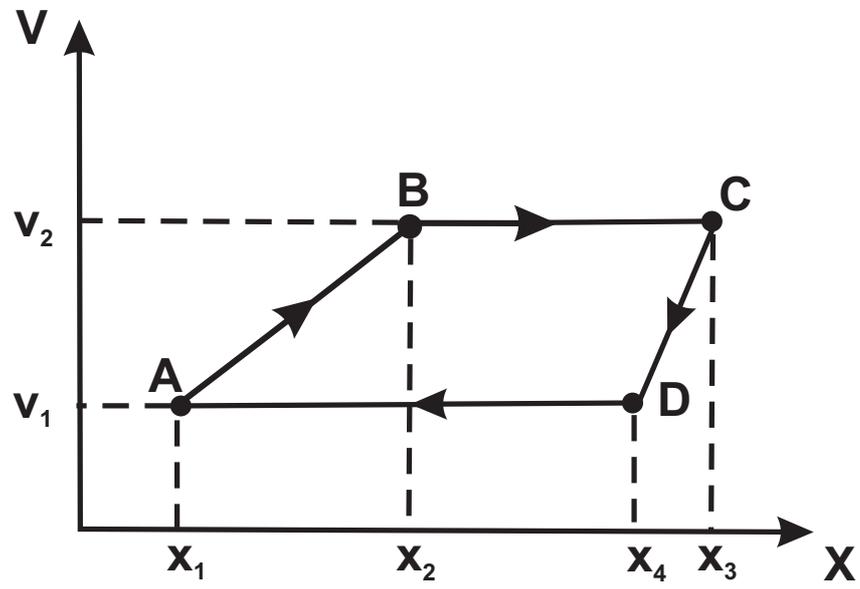

Figure 1

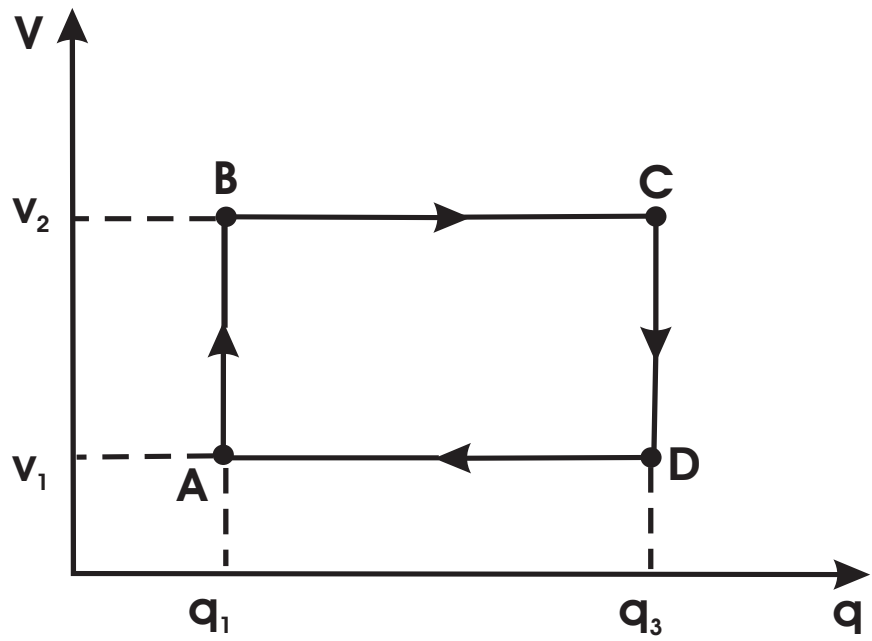

Figure 2